\documentclass[aps,prl,twocolumn,footinbib,superscriptaddress]{revtex4-1}
\usepackage{amsmath,amssymb,amsfonts}
\usepackage{bm}
\usepackage{color}
\usepackage{graphicx}
\usepackage{float}
\usepackage{wrapfig}
\usepackage{verbatim}
\usepackage{gensymb} 
\usepackage[english]{babel}
\usepackage[colorlinks=true,bookmarks=false,citecolor=black,urlcolor=black,,linkcolor=black]{hyperref} 
\usepackage{natbib}

\usepackage{subfigure}
\usepackage{amstext, mathrsfs, textcomp}
\usepackage{multirow}
\usepackage{xcolor}
\usepackage{enumerate}
\usepackage{epstopdf}
\usepackage{tabularx}
\makeatletter

\newcommand{\ket}[1]{\left|#1\right>}      

\begin{document}

\title{Third-harmonic generation in photonic topological metasurfaces}

\author{Daria Smirnova}
\affiliation{Nonlinear Physics Centre, Australian National University, Canberra ACT 2601, Australia}

\author{Sergey Kruk}
\affiliation{Nonlinear Physics Centre, Australian National University, Canberra ACT 2601, Australia}

\author{Daniel Leykam}
\affiliation{Center for Theoretical Physics of Complex Systems, Institute for Basic Science (IBS), Daejeon 34126, Republic of Korea}

\author{\\Elizaveta Melik-Gaykazyan}
\affiliation{Nonlinear Physics Centre, Australian National University, Canberra ACT 2601, Australia}
\affiliation{Faculty of Physics, Lomonosov Moscow State University, Moscow 119991, Russia}

\author{Duk-Yong Choi}
\affiliation{Laser Physics Centre, Australian National University, Canberra, ACT 2601, Australia}

\author{Yuri Kivshar}
\affiliation{Nonlinear Physics Centre, Australian National University, Canberra ACT 2601, Australia}



\begin{abstract}
We study nonlinear effects in two-dimensional photonic metasurfaces supporting topologically-protected helical edge states at the nanoscale. We observe strong third-harmonic generation mediated by optical nonlinearities boosted by multipolar Mie resonances of silicon nanoparticles. Variation of the pump-beam wavelength enables independent high-contrast imaging of either bulk modes or spin-momentum-locked edge states. We demonstrate topology-driven tunable localization of the generated harmonic fields and map the pseudospin-dependent unidirectional waveguiding of the edge states bypassing sharp corners. Our observations establish dielectric metasurfaces as a promising platform for the robust generation and transport of photons in topological photonic nanostructures.
\end{abstract}

\maketitle

Topological photonics describes optical  structures with the properties analogous to electronic topological insulators~\cite{Lu2016}. These systems are distinguished by bulk band gaps that host disorder-robust states localized at edges or interfaces and provide a novel approach for designing non-reciprocal or localized modes for optical isolators, photonic-crystal waveguides, and lasers. Since the original demonstration of backscattering-immune photonic topological edge states with the use of a gyrotropic microwave photonic crystal under a strong magnetic field~\cite{Wang2009}, there has been a concerted effort towards realizing topological photonics at the nanoscale. Recently suggested optical designs compatible with non-magnetic all-dielectric structures~\cite{Wu2015,Slobozhanyuk2016,Ma2016,Barik2016,Anderson2017,Shalaev2018NNANO} are now emerging as a promising platform for quantum and nonlinear topological photonics~\cite{Bahari2017,Segev2018b,Barik2018,Mittal2018}.

Stimulated by the progress in nanofabrication techniques, a new favourable ground for topological photonics based on dielectric nanoparticles with high refractive index has recently emerged~\cite{Kuznetsov2016,Rider2019}. Strong optical resonances and low Ohmic losses make it feasible for practical implementation of topological order for light at subwavelength scales. 
The underlying conceptual framework is to use arrays of meta-atoms with judiciously engineered shape and lattice structure, with topologically nontrivial features arising from 
pseudospin degrees of freedom. It bridges fundamental physics of topological phases with resonant nanophotonics and multipolar electrodynamics~\cite{Smirnova2016,Smirnova:16}. 
Topological metasurfaces could form a ground for a new class of ultra-thin devices with functionalities based on novel physical  principles through engineering light-matter interactions
in synthetic photonic potentials~\cite{Ni2018}. 
Their pseudospin-dependent physics may be useful for manipulation of internal degrees of freedom of light such as polarization and angular momentum. 

However, the experimental characterization of topological photonic structures becomes much more challenging at the nanoscale.
Most implementations so far have been limited to \emph{indirect probing} of topological states such as transmission spectra~\cite{Barik2018,Gorlach2018,Shalaev2018,He2018}, which cannot provide spatially-resolved information about the edge modes and suffer from input/output coupling losses. Other recently-demonstrated linear approaches such as near-field imaging~\cite{Kruk2017zigzag}, cathodoluminescence~\cite{Peng2019}, and far-field imaging~\cite{Parappurath2018arxiv} suffer from poor spatial resolution or small field of view,  leaving the edge states almost completely hidden in the background noise. Direct high-contrast imaging of the edge states is essential for assessing the fidelity of the topological waveguides, identifying potential sources of backscattering, and for optimizing the coupling between the edge states and localized emitters~\cite{Ma2016}.

Here we show that nonlinear topological photonics provides an effective way to overcome these limitations by observing nonlinear light conversion in a topological photonic nanostructure. By varying the frequency and polarization of a pump beam and measuring the generated third-harmonic signal, we demonstrate selective imaging of either bulk modes or edge modes that are otherwise undetectable via conventional linear far field imaging. Our approach significantly enhances the measurement contrast, sensitivity, and imaging area compared to other recent works, enabling us to unambiguously visualize nanoscale topological photonic edge states and their propagation around corners and defects. Importantly, our platform combining appreciably strong optical nonlinearity with topological band structures paves the way towards observing other nonlinear wave interactions in photonic topological insulators.

\begin{figure} [t!]
\centering
\includegraphics[width=0.99\linewidth]{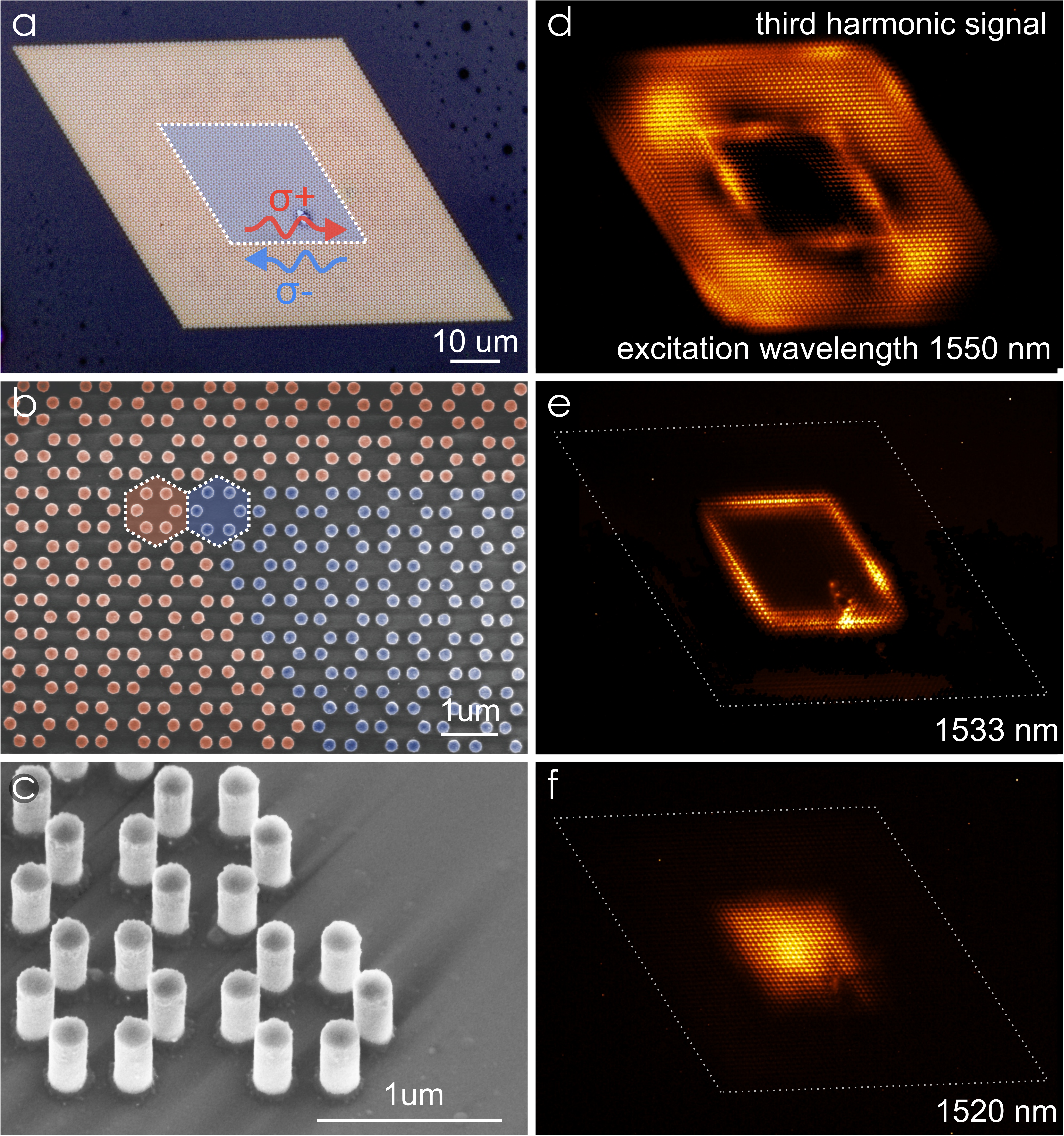}
\caption{Experimental results for dielectric metasurfaces with topological edge states. (a) Optical microscope image of a topological metasurface guiding robust edge waves with opposite helicities ($\sigma^{\pm}$) along the interface between topologically different outer and inner domains (painted blue). Dashed line -- guide for a eye.
(b,c) Scanning electron microscopy images: (b) top view and (c) side view of the metasurface consisting of silicon pillars arranged into hexagon clusters. (b) domain wall between expanded (blue) and shrunken (red) domains. Framed hexagons highlight corresponding unit cells.
Pillars' radius $r=105$~nm, height $h=538$~nm, lattice constant of hexagon clusters $a=1100$~nm,  shrink/expand coefficients are $0.95$ and $1.05$, correspondingly.  
(d-f) Experimental images of third-harmonic intensity distribution for three excitation wavelengths.}
 \label{fig:fig1}
\end{figure}

\begin{figure*} [t!]
\centering 
\includegraphics[width=0.99\linewidth]{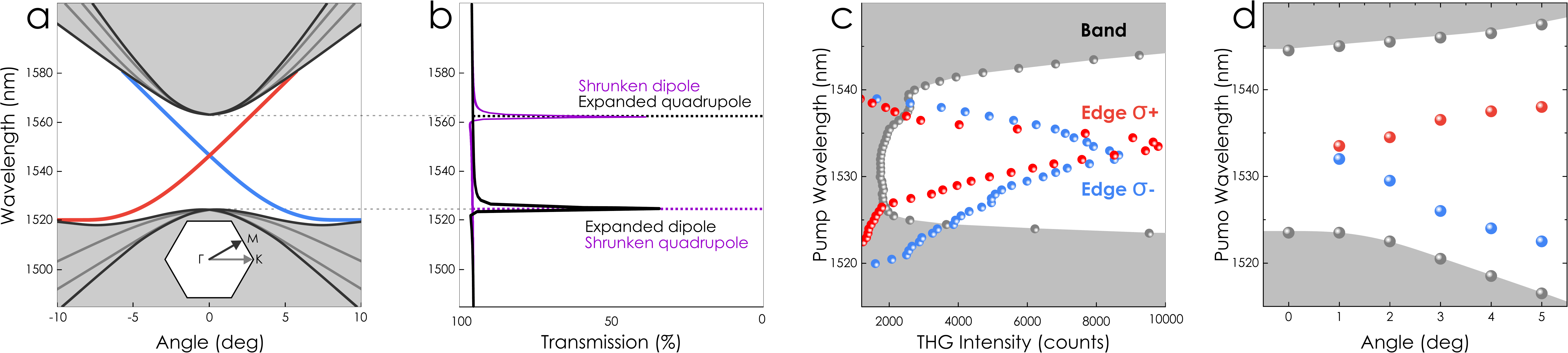}
 \caption{ 
Dispersion properties of the topological helical edge states. (a) Numerically calculated photonic band structures for the four doublet bands along $\Gamma-M$ (black curves) and $\Gamma-K$ (dark gray curves) directions, 
found almost perfectly overlaid for shrunken and expanded arrays. The bulk bands are shaded in light gray, while the dispersion branches of the edge modes spanning the topological band gap are shown by colored curves. Inset: Brillouin zone. (b) Simulated transmission spectra of the infinite expanded and shrunken metasurfaces featuring excitation of dipolar resonances at normal incidence. Note that the optical response of the fabricated sample is spectrally blue-shifted by approximately 15 nm. (c) Experimentally measured TH spectrum indicating the mid-gap edge states excited at the interfaces of the rhomboid topological cavity. (d) Spectral positions of TH maxima depending on angle of incidence trace diverging dispersion branches for opposite circular polarizations.
 }
 \label{fig:fig2}
\end{figure*}

For our experiments, we design photonic all-dielectric metasurfaces exhibiting a topological phase transition and band inversion above the light line, similar to an earlier theoretical proposal~\cite{Wu2015}. The array of silicon nanopillars hexamers is divided into two domains -- expanded or shrunken hexamers -- characterized by distinct topological invariants. Their interface supports a pair of topologically protected, spin-momentum locked edge states~\cite{Wu2015,Yves2017,Gorlach2018,noh2018topological}.

We fabricate our samples from silicon on a glass substrate [for details, see Supplemental Material]. 
Figures~\ref{fig:fig1}(a-c) show optical and  electron microscope images of one of the fabricated samples, consisting of a rhomboid-shaped domain of expanded hexamers embedded in a domain of shrunken hexamers. We illuminate the sample with a powerful short-pulse laser with a tunable wavelength~\cite{Wang2018} [see details in 
Supplemental Material]. The laser beam size is larger than the total size of the sample. The strong cubic nonlinearity of silicon~\cite{Gai2014,Shorokhov2016} naturally provides optical frequency conversion capabilities, and the resonant near-field enhancement provided either by bulk or edge modes boosts the nonlinear harmonic generation~\cite{Smirnova:16}, enabling accurate mapping of the corresponding modes. The generated third-harmonic (TH) radiation is imaged on a camera [see
Supplemental Material]. 

Figures~\ref{fig:fig1}(d-f) show three representative cases of the TH distribution at different pump frequencies: in Figs.~\ref{fig:fig1}(d,f) the TH comes from the bulk of the shrunken/expanded regions. This corresponds to resonant excitation of bulk dipolar modes in these two domains. In Fig.~\ref{fig:fig1}(e) the pump is tuned to the bulk band gap of the two domains, and the TH signal is generated along the domain wall between the shrunken and expanded structures, visualizing the topological edge states. We notice that in our experiments the coupling to the edge states at normal incidence is inefficient, and the result presented in Fig.~\ref{fig:fig1}(e) is an average of two images obtained at $\pm$1$\degree$ incident angles. To demonstrate the potential of the nonlinear diagnostics technique, we additionally perform sets of complimentary \emph{linear measurements} for identical experimental configurations with the results presented in Supplemental Material. With linear approach we are able to observe convincingly the excitation of bulk modes, although with a substantially degraded signal-to-noise ratio. However, the topological light localization along the domain wall remains invisible in our linear experiments. It is vastly easier to observe edge states using the conventional linear imaging in the large-scale designs of topological systems based on the waveguide geometry~\cite{Noh2018}, with a typical size of the building blocks much larger than the wavelength of light. But this platform is incompatible with planar silicon-based integrated optics.

To explain the wavelength-dependent optical response and the observed edges states, we model each nanopillar as an out-of-plane dipole with predominant $E_{z}$ component of the electric field directed along the vertical axis. The collective modes supported by the individual hexamers can be classified according to their in-plane symmetry, with the bands in the spectral range of interest composed primarily of dipole ($p$) and quadrupole ($d$) hexamer eigenmodes. In the vicinity of the $\Gamma$ point, the bulk photonic band structure is captured by the eigenvalue problem $\mathcal{E}\,\ket{\psi_{\pm}}=\hat{H}_{\pm}\,\ket{\psi_{\pm}}$,
with the effective Hamiltonian forming $2\times 2$ blocks $\hat{H}_{\pm}$ given by 
\begin{equation}\label{BlockForm}
\hat{H}_{\pm}=\begin{pmatrix}
\mu+\beta\,k^2 & v\,(\mp k_x-i\,k_y) \\
v\,(\mp k_x+i\,k_y) & -\mu-\beta\,k^2
\end{pmatrix}\:,
\end{equation}
corresponding to the two circular polarizations of eigenstates $\ket{\psi_{\pm}}=\left(\ket{p_{\pm}},\ket{d_{\pm}}\right)^T$~\cite{Gorlach2018}. Equation~\eqref{BlockForm} incorporates effective parameters: $\mu$ and $\beta$ are the mass term and band parabolicity, respectively, $v$ is velocity; $k_{x,y}$ are in-plane components of the wavevector. The two decoupled polarizations form a pseudospin degree of freedom and thus the spin Chern number $C=\left(C_{-}-C_{+}\right)/2$ can be introduced, where $C_{\pm}$ is Chern number for each individual block~\cite{Shen}. Each of the blocks Eq.~\eqref{BlockForm} has the structure of a Dirac-like Hamiltonian, for which the spin Chern number can be straightforwardly calculated as $C=\frac{1}{2}\,(\text{sgn}\,\mu-\text{sgn}\,\beta)\:$~\cite{Shen}. The shrunken domain is described by the condition $\mu \beta >0 $ ($C=0$; trivial), whereas the condition $\mu \beta < 0 $ is valid for the expanded domain ($C=1$; nontrivial), in accordance with Ref.~\cite{Wu2015}; thus the interface states are topologically protected. The edge states bound to the interface $x=0$ exhibit linear dispersion crossing $\mathcal{E}_{\text{e.s.}} (k_y) = \pm v k_y$.

Figure~\ref{fig:fig2}(a) shows the numerically computed bulk band diagram of the structure [for details of
numerical simulations, see Supplemental Material]. The bulk modes are primarily quadrupolar or dipolar, and the transition from shrunken to expanded designs is accompanied by a band inversion. At the normal incidence, the external radiation can only couple to the dipolar modes, as seen in the calculated transmission spectra in Fig.~\ref{fig:fig2}(b). When the pump is tuned to the upper band edge, it is resonant only with the dipolar bulk modes of the shrunken outer domain, whereas at the lower band edge it is resonant with the dipolar bulk modes of the expanded inner domain. Hence, the localization of nonlinear light generation occurs, as the wavelength is varied mapping the band inversion. 

Figure~\ref{fig:fig2}(a) additionally illustrates the characteristic Dirac-like dispersion of the spin-momentum locked edge states residing in the band gap, which can be selectively excited using a circularly polarized pump beam. Figure~\ref{fig:fig2}(c) shows the experimental TH spectra for 1$\degree$ incidence measured from the bulk (gray dots) and from the domain wall excited with the two orthogonal circular polarizations of the pump laser beam (red and blue). The two polarizations excite the edge modes with the opposite helicity values $\sigma^{+}$ and $\sigma^{-}$.

\begin{figure} [b!]
\centering
\includegraphics[width=0.99\linewidth]{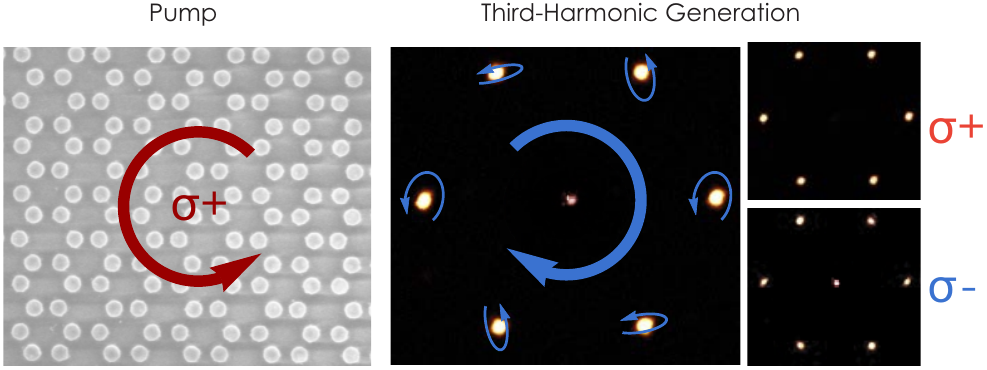}
\caption{Nonlinear diffraction and polarization conversion.
Experimentally observed back-focal plane images of the TH signal generated by an all-dielectric metasurface consisted of an inner domain of shrunken hexamers and outer domain of expanded hexamers in the forward direction for the left-handed circularly polarized pump, $\sigma^{+}$, at $0^{\circ}$ incident angle for the fundamental wavelength of 1540~nm which corresponds to a bulk state. Blue arrows visualize the polarization state of the TH fields. Right: back-focal plane patterns of the TH field for two states of an analyzer: left-handed (top) and right-handed (bottom) circular polarized.}
 \label{fig:TH_Polarimetry}
\end{figure}

We experimentally measure the dispersion of the edge states by tilting the sample along its horizontal axis between 0$\degree$ and 5$\degree$ and tracing the maxima of the TH spectra [see Fig.~\ref{fig:fig2}(d)]. Since our domain wall forms a closed cavity, to distinguish between the $\sigma^{+}$ and $\sigma^{-}$ states we sample the TH intensity from the upper part of the domain wall only (on the lower edge their dispersion is opposite). Figure~\ref{fig:fig2}(d) shows the spectral separation of the $\sigma^{+}$ and $\sigma^{-}$ states as the angle increases. In particular, their dispersion is close to linear, indicating the edge states are nearly gapless and decoupled, consistent with their topological origin. 

Nonlinear diffraction from the metasurface sample exhibits a characteristic hexagonal far-field pattern. It is accompanied by the circular polarization conversion due to the nonlinear interference of the induced multipolar sources related to the excited photonic modes of the metasurface [see details in {Supplemental Material}]. 
We perform the polarimetry of the third-harmonic signal and observe the polarization conversion of $\sigma^{+}$ pump into $\sigma^{-}$ harmonic and vice verse in the zeroth diffraction order, as summarized in Fig.~\ref{fig:TH_Polarimetry}. The nonlinear conversion efficiency is estimated to be comparable to the efficiency of previously demonstrated silicon Mie-resonant metasurfaces~\cite{Smirnova:16,kruk2017functional} [see details in {Supplemental Material}].

\begin{figure} [b!]
\centering
\includegraphics[width=0.99\linewidth]{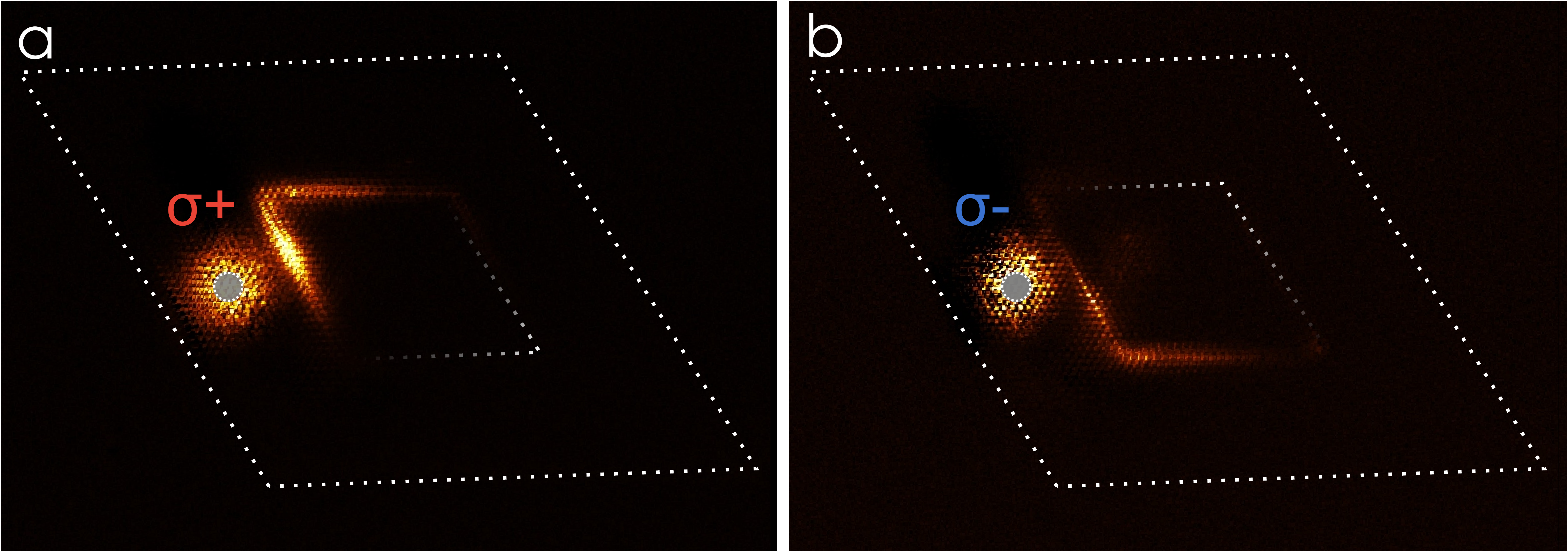}
\caption{Spin-momentum-locked topological guided modes. Experimental images of the third-harmonic intensity distribution for the pump focused within the dashed circle with (a) left-circular, and (b) right-circular polarizations. }
 \label{fig:SH}
\end{figure}

Next, we study the spin-momentum-locked waveguiding of the optical edge modes associated with the analogue of the quantum spin Hall effect for light~\cite{Bernevig2006,Wu2015,Bliokh2015,bliokh2015spin}. In this design the edge states are topologically protected by the $C_{6v}$ rotational symmetry of the hexamers~\cite{Wu2015}. As long as this symmetry is preserved, the counter-propagating edge states remain decoupled and cannot scatter backward, even when the interface has sharp corners, as in our system. To excite and map the counter-propagating edge modes individually, we locally focus the circularly polarized pump beam in a spot near the domain wall [see the dashed circle in Figs.~\ref{fig:SH}(a,b)]. In Fig.~\ref{fig:SH}(a) the pump with $\sigma^{+}$ polarization
couples to the mode propagating clockwise, while in Fig.~\ref{fig:SH}(b) the $\sigma^{-}$ pump
launches  the counter-clockwise wave propagation. Both waves propagate along the domain wall passing the corners.

\begin{figure}[t!]
\centering
\includegraphics[width=0.99\linewidth]{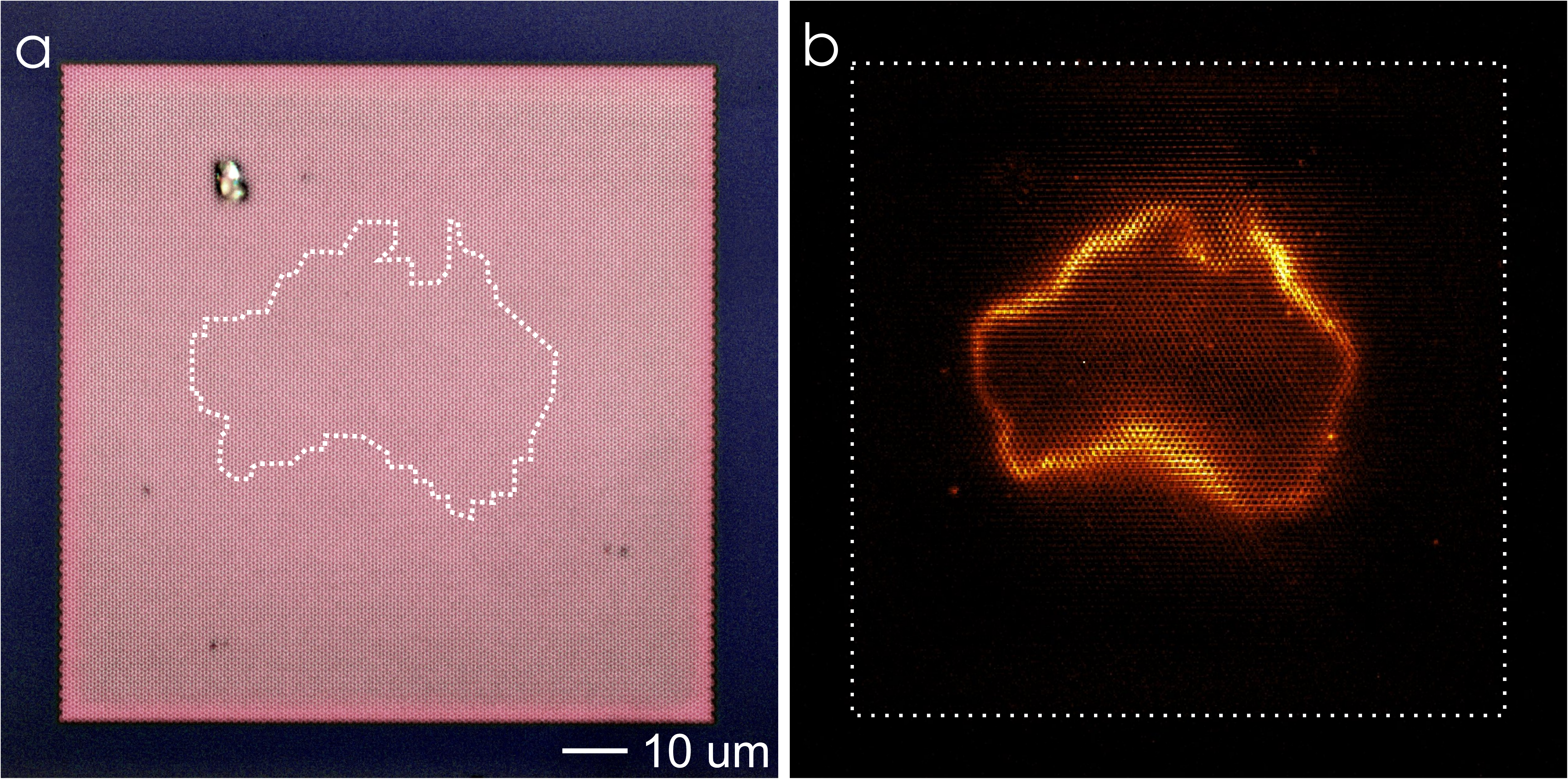}
\caption{Topological Australia: An example of a third-harmonic image generated from geometry-independent edge states. (a) Optical microscope image of a metasurface with a domain wall of a shape of the Australian continent. Dashed line is a guide for an eye for the domain wall between the shrunken and the expanded domains. (b) Experimentally observed third-harmonic field at the edge-state pump wavelength.}
 \label{fig:AUS}
\end{figure}

Finally, we demonstrate the existence of the edge states for arbitrary geometries of the topological interfaces. For this, we fabricate a metasurface with a domain wall similar to the shape of the Australian continent, see Fig.~\ref{fig:AUS}. When the whole metasurface is pumped with the wavelength corresponding to that of the edge mode, the domain wall becomes clearly imaged via the third-harmonic field contour.

In summary, we have suggested and demonstrated a novel method to image topological edge states at optical frequencies using third-harmonic generation. Compared to conventional linear imaging methods, our nonlinear approach offers superior contrast, sensitivity, and imaging area, enabling both characterization and optimization of topological waveguides for quantum and nonlinear optics applications. We have observed pseudospin-momentum locking of edge photonic modes at the topological interfaces, verifying their ability to propagate around sharp corners without backscattering. Furthermore, we have demonstrated tunable localization and enhancement of harmonic generation in nanoscale topological photonic structures. 
Our results make a step towards bridging nonlinear optics with topological physics for the integrated and robust generation and transport of photons at the nanoscale.

\begin{acknowledgements}
The authors thank B.~Luther-Davies for his support with tunable lasers, M. Lockrey for his help with the electron microscopy, and K.~Bliokh for useful discussions. They acknowledge the use of the Australian National Fabrication Facility of the ACT Node. The work has been supported by the Strategic Fund of the Australian National University, the US Air Force Office of Scientific Research (grant FA2386-16-1-0002), and the Institute for Basic Science (IBS-R024-Y1). 
\end{acknowledgements}



\end{document}